# Analysis of the structural characteristics and optoelectronic properties of CaTiO$_3$ as a non-toxic raw material for solar cells: a DFT study

Nematov D.D., Burhonzoda A.S., Shokir F.

*S.U.Umarov Physical -Technical Institute of the National Academy of Sciences of Tajikistan*

**Abstract:** Structural and optoelectronic properties of α, β, γ phases of calcium titanate are studied with the implementation of first-principles quantum-chemical calculations in the framework of DFT. When optimizing the geometry, the GGA approximation was used. The relaxed lattice parameters obtained by us are identical with the experimental analogs. It has been established that the most stable phase of calcium titanate is the orthorhombic syngony, which corresponds to the results of experimental measurements. The optoelectronic properties of these materials have been studied using the Wien2k code. The high-precision TB-mBJ approximation was used to calculate the exchange-correlation effects. An analysis of the electronic properties of these materials showed that all the studied phases of calcium titanate belong to the class of wide-gap semiconductors. The calculated band gaps for the cubic, tetragonal, and orthorhombic CaTiO$_3$ systems are 2.83, 3.07, and 3.26 eV, respectively. According to the analysis of DOS-plots, it was found that the tetragonal phase of calcium titanate is characterized by the highest density of states. Calculations of the optical constants of the systems under study showed that the CaTiO$_3$ cubic system is characterized by an increased absorption capacity and a relatively high photoconductivity. However, for the other two phases of calcium titanate, the calculations gave identical patterns, i.e., the absorption and optical conductivity spectra of the tetragonal and orthorhombic CaTiO$_3$ systems practically coincide.

## 1. Introduction

In the process of human activity, technology and industry develop and improve various areas of people's lives on earth, but along with the steady development of science and technology, such global problems as environmental pollution and climate warming are exacerbated day by day [1]. On the other hand, the world's non-renewable energy resources, such as hydrocarbon fuels, are being depleted at a high rate. Obviously, these problems pose a great threat to humanity and the development of life on Earth, and failure to take the necessary measures in the future may make life on the planet difficult or even impossible. In this regard, humanity needs to find alternative ways to solve these problems and how to quickly prevent the accelerated destruction of energy resources and global warming, as well as the melting of large glaciers.

One of the most effective ways to combat global warming is to replace fossil fuels with new energy-efficient materials and various renewable energy sources [1]. Based on this, the developed countries of the world, especially the countries of Europe, from year to year increase their investments in this area, so that scientists and engineers of the world as soon as possible develop modern means and new weapons to combat atmospheric pollution and achieve sustainable development of technologies and green energy [2]. The very idea of developing new energy-efficient materials and switching to renewable energy sources is in line with the UN strategy to prevent global problems (paragraphs 7 and 13 of the Sustainable Development Strategy) for the period up to 2030 [3].

There are several viable renewable energy sources, among which photovoltaic generators are considered the most promising method of replacing fossil fuels. Photovoltaic generators are based on semiconductor materials, which in recent years have been obtained from polycrystalline and monocrystalline silicon compounds. Thanks to this, the efficiency of silicon solar cells has reached 26.1% in recent years [4], but despite this, there are a number of problems, such as complex methods for the synthesis and processing of crystals, the presence of photochemical degradation and a relatively high price, which obliges scientists and researchers around the world to develop new, more advanced, cheap and highly efficient materials.

In the process of developing optimal materials, scientists and engineers have found another alternative way out - to use the unique properties of calcium titanate [5]. This is how tandem photovoltaic cells were born, in which the perovskite layer worked in parallel with the electroactive silicon layer. In addition, calcium titanate is cheaper to produce commercially and easier to manufacture. However, in the beginning, scientists did not pay enough attention to these compounds and were engaged in the search and development of other materials. Further, new generation perovskite solar cells have been developed and spread very rapidly in the past few years due to the outstanding photovoltaic properties of the organo-inorganic perovskite layer and the exceptional efforts of scientists and engineers to research and improve their properties [6]. For example, in 2009, research by Miyasaki and others used hybrid lead perovskite $CH_3NH_3PbI_3$ ($MAPbI_3$) as light absorbers in solar cells and demonstrated an energy efficiency of 3.8%. [7]. Over the years, scientists have proposed other varieties of lead organic-inorganic perovskites, as well as their flexible panels, which have been put into large-scale operation day after day. However, the efficiency share of these hybrid-organo-inorganic perovskites rapidly increased from 3.8% [8] to 25.7% [4] from 2009 to 2021, rivaling silicon solar cells in efficiency [9]. On the other hand, despite the high efficiency of solar panels, these materials have the problem of stability and toxicity due to the presence of lead in their composition, which limited the further step of their commercialization [10]. Then, in search of an alternative, the researchers again returned to the old and well-known natural perovskite based on calcium titanate.

Calcium titanate ($CaTiO_3$) is one of the wide-gap semiconductor perovskites (the only natural, but ineffective) with the general formula $ABX_3$, where the A cation occupies a cuboctahedral position, and the B cation octahedral. X is a halide anion [11]. In addition to semiconductor properties, $CaTiO_3$ also exhibits dielectric properties with a relative permittivity of up to 186 and a band gap of 3–4 eV, which can be used as an optoelectronic device [12]. Depending on the phase transition temperature, $CaTiO_3$ can be divided into four space groups: orthorhombic (Pbnm), orthorhombic (Cmcm), tetragonal (I4/mcm), and cubic (Pm3m). Among them, the cubic phase is formed at a high temperature (T > 1300°C), and the tetragonal phase is a transition compound that can only form at a very limited temperature (1250°C < T < 1349°C). The orthorhombic phase (Pbnm) is stable at room temperature [13].

Interest in this mineral, as a potential semiconductor for photovoltaic systems, arose only in the 21st century, with the advent of thin-film technologies. The very first experiments confirmed that perovskite solar cells carry out the transfer of electric charge no worse than the "classics" of silicon. But in this case, the specific absorption of the same amount of radiation was achieved at a thickness of 180 μm for a silicon wafer and 1 μm for a perovskite film. The reason for this turned out to be approximately the same times greater effective width of the absorption spectrum of an inconspicuous mineral, that is, the absorption range of rays and the photoelectric activity of calcium titanate are much wider than those of other perovskite structures. It is obvious that such wide-gap semiconductors are able to work without problems under conditions of intense UV radiation and/or simultaneously become a dye for solar cells in the IR range. On the other hand, metal doping can reduce the band gap of $CaTiO_3$ and optimize its optical properties for direct conversion of IR radiation. However, mountainous countries with a special climate, such as Tajikistan, are characterized by intense solar radiation and high light radiation. Therefore, despite the fact that Tajikistan has huge hydropower resources and the demand for electricity is rather low, the creation of a photovoltaic solar power plant is one way to meet the demand for electricity in high mountainous remote areas and develop green energy in this country.

In this work, we study the structural properties, electronic structure and optical spectra of the α, β, γ phases of $CaTiO_3$ through theoretical quantum chemical calculations in the framework of DFT, in order to understand the absorption features of a wide spectrum of light based on them.

## 2. Computing details

Ab initio DFT calculations to study the electronic properties of the α (Pm3mm), β (I4/mcm), γ (Pbnm) $CaTiO_3$ phases were performed using the Wien2k package [14] after GGA relaxation of all structures under study in the VASP code [15]. The electronic and ionic relaxation of the structures under study was achieved at a cutoff energy of 550 eV and a uniform grid of 2x2x2 k-points. The structures relaxed until the numerical convergence in self-consistent cycles reached forces between ions less than $10^{-3}$ eV Å$^{-1}$, and the total energy of the system did not reach $10^{-4}$ eV. When optimizing the structures under study, the symmetry was maintained constant, and the position of the atoms, the volume, and the shape of the cell were allowed to relax. The optical properties, electronic structure, and band gap of the systems under study were estimated using the TB-mBJ exchange-correlation function. The Wien2k calculations used the 1000 k − point in the Brillouin zone for self-consistent convergence with a cutoff RMT∗Kmax = 7, where RMT is the smallest radius of the muffin ("atomic") sphere and Kmax is the magnitude of the largest wave vector. All computer calculations were performed on a high-performance computing cluster equipped with 1 computing node with one physical processor Intel Core I9-9960X CPU 3.10 G (16 cores at 3.0 GHz) and 32 GB of RAM.

### 3. Results and Discussion

At the first stage of the research, geometry optimization was carried out, from which the relaxation parameters of the cell of the systems under study were obtained. Relaxed structures of $CaTiO_3$ in α-, β- and γ-phases are shown in Figure 1.

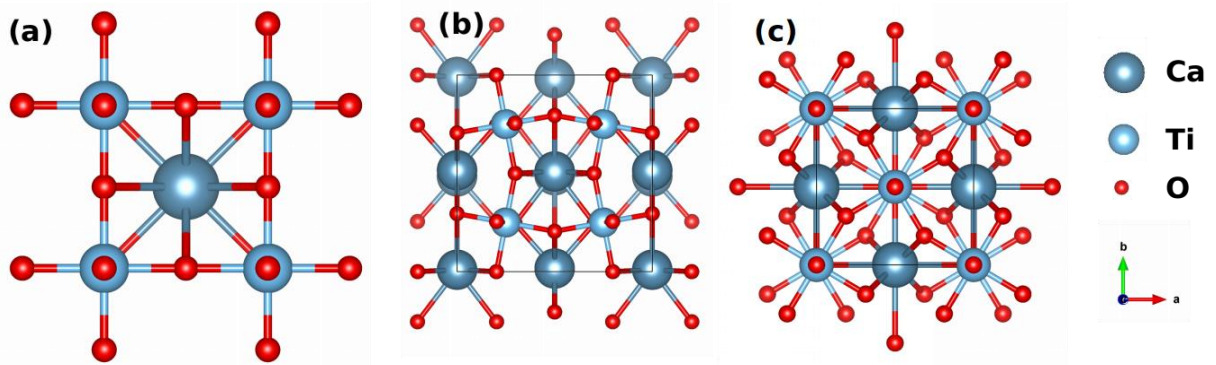

**Figure 1.** Relaxed structures of the α (a), β (b) and γ (c) phases of $CaTiO_3$

Further, the calculated parameters and lattice volume of the α-, β- and γ-phases of perovskite $CaTiO_3$ were calculated and are given in Table. 1. It can be seen that the results obtained are in good agreement with other results of other independent researchers [17–18]. According to the results of calculating the minimum energy (energy of the ground state) of the studied phases, it can be seen that the most stable structure of calcium titanate is the orthorhombic system (Pbnm) of this compound (followed by the tetragonal structure I4/mcm and the cubic structure Pm3m), which confirms the results of theoretical MD calculations [19] and experimental measurements [20-21].

**Table 1.** Comparison of calculated and experimental values of lattice constants and ground state energy $\alpha$, $\beta$, and $\gamma$ of CaTiO$_3$

| Lattice parameters | Phase | | | | | |
|---|---|---|---|---|---|---|
| | $\alpha$ | | $\beta$ | | $\gamma$ | |
| | This work | Other works | This work | Other works | This work | Other works |
| **a (Å)** | 3.8574 | 3.896[18] | 7.6032 | - | 5.3548 | 5.379[17] 5.498[18] |
| **b (Å)** | 3.8574 | 3.896[18] | 7.6888 | - | 5.3548 | 5.442[17] 5.498[18] |
| **c (Å )** | 3.8574 | 3.896[18] | 7.6864 | - | 7.8203 | 7.640[17] 7.78[18] |
| **V (Å$^3$)** | 57.39 | - | 449.34 | - | 224.247 | 223.6[17] |
| **E$_0$ (eV)** | -40.04 | - | -159.91 | - | -319.97 | - |

The relaxed positions of atoms and their smallest radii in the crystal structure of the $\alpha$, $\beta$, and $\gamma$ phases of CaTiO$_3$ perovskite are given in Table 2.

**Table 2.** Atomic coordinates in $\alpha$, $\beta$, and $\gamma$ phases of perovskite CaTiO$_3$.

| Phase | Atom type | x | y | z | RMT |
|---|---|---|---|---|---|
| $\alpha$ | Ca1 | 0.00 | 0.00 | 0.00 | 2.50 |
| | Ti1 | 0.50 | 0.50 | 0.50 | 1.90 |
| | O1 | 0.50 | 0.00 | 0.50 | 1.72 |
| $\beta$ | Ca1 | 0.00 | 0.50 | 0.75 | 2.17 |
| | Ca2 | 0.00 | 0.98 | 0.25 | 2.17 |
| | Ti1 | 0.25 | 0.75 | 0.00 | 1.93 |
| | O1 | 0.00 | 0.79 | 0.54 | 1.74 |
| | O2 | 0.20 | 0.75 | 0.25 | 1.74 |
| | O3 | 0.20 | 0.50 | 0.00 | 1.74 |
| $\gamma$ | Ca1 | 0.50 | 0.00 | 0.75 | 2.27 |
| | Ti1 | 0.00 | 0.00 | 0.00 | 1.93 |
| | O1 | 0.31 | 0.18 | 0.50 | 1.74 |
| | O2 | 0.00 | 0.00 | 0.25 | 1.74 |

Interatomic distances in the studied phases of CaTiO$_3$ perovskite were calculated and presented in Table 3.

**Table 3.** Distance between atoms in α, β and γ phases of perovskite CaTiO$_3$

| Phase | Bond | Bond length (Å) |
|-------|------|-----------------|
| α | Ca-O | 2.72 |
|   | Ti-O | 1.92 |
| β | Ca-O | 2.31 |
|   | Ti-O | 1.95 |
| γ | Ca-O | 2.67 |
|   | Ti-O | 1.95 |

According to the results presented in tables 1-3, during the transition from one phase to another phase of CaTiO$_3$, the distance between cations and anions in the system increases, which leads to the volume expansion of their lattices. Nevertheless, Fig. 2 shows X-ray patterns of the studied samples, from which it is easy to conclude that during the structural-phase transition of calcium titanate, the interplanar distance and, accordingly, the lattice parameters in it change.

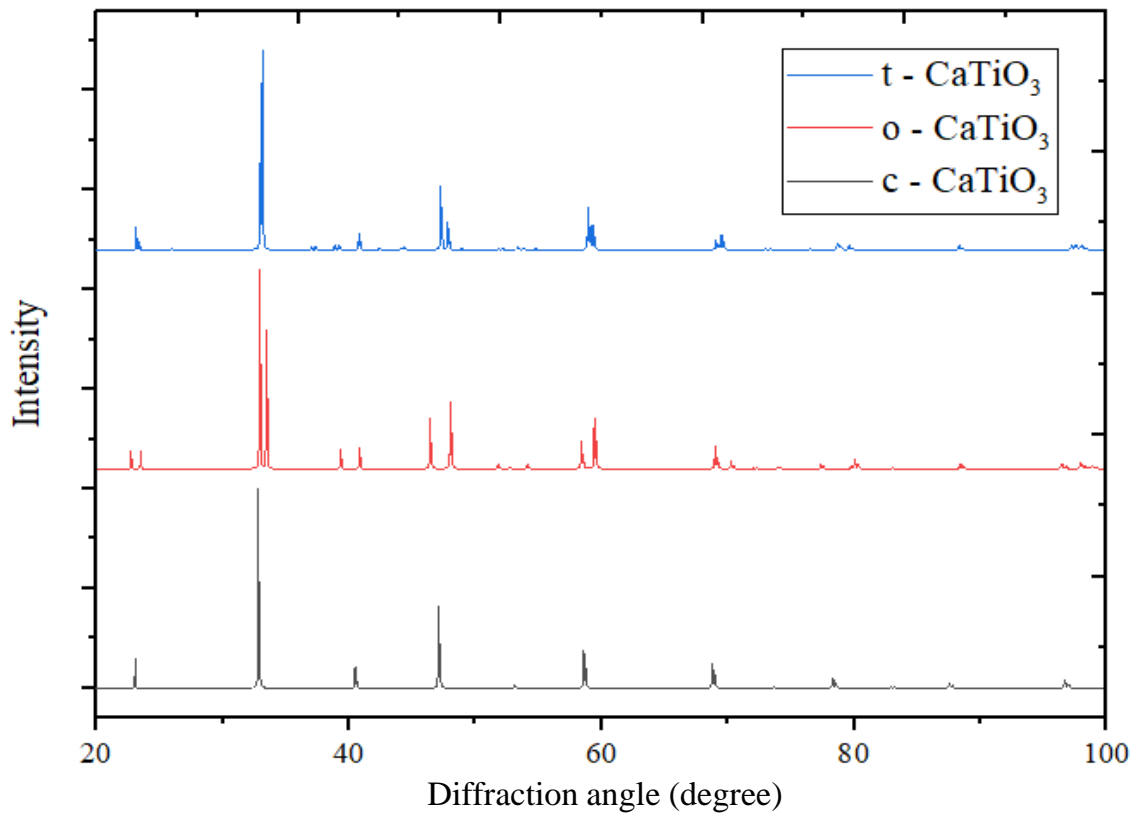

**Figure 2.** Dependence of the intensity on the diffraction angle for the α, β and γ phases of perovskite CaTiO$_3$

This behavior of the system is further carried out to change the absorptivity and band gap of the studied phases of calcium titanate. It can also be concluded that the result of changing the interatomic distance and interplanar distance leads to an increase in the density of electronic states and the formation of vacancies in the atomic orbitals of the final phases of calcium titanate. It can be seen from the results that during the α-γ structural phase transition, the X-ray peaks are mixed with increased intensity towards larger angles. An increase in the half-width of X-ray peaks is also observed.

The density of state (DOS) is one of the most important properties that provides information about the electronic nature of all solid-state systems. DOS also allows you to determine the nature of the chemical bonds between the constituent atoms. In this work, the total density of states (DOS) of cubic, tetragonal, and orthorhombic phases of calcium titanate after the final relaxation of their crystal structures by the GGA potential was studied by the TB-mBJ method. The results of calculations of the DOS of the studied materials depending on the energy in the range of -5 eV and 5 eV are shown in Fig. 3. The Fermi level is given at 0 eV. The results show that all studied $CaTiO_3$ phases are semiconductors with a wide bandgap.

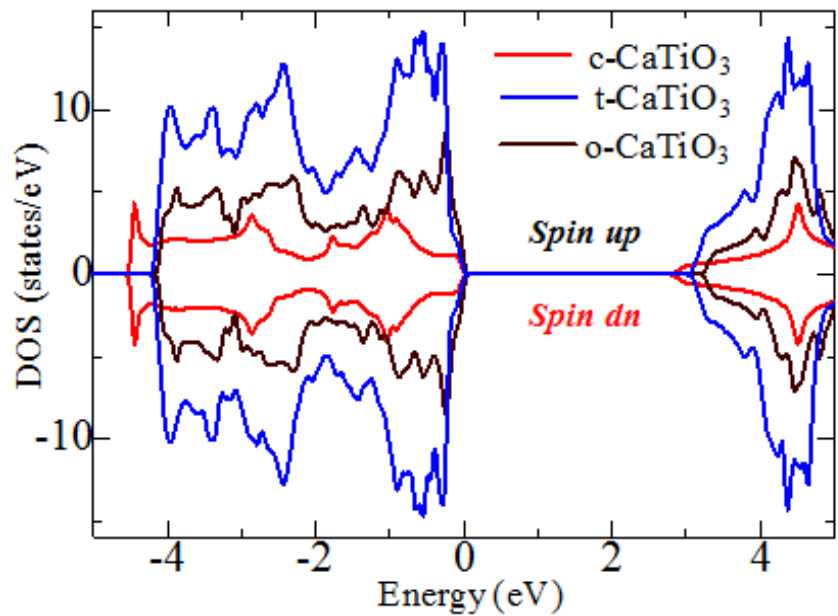

**Figure 3.** DOS of α, β, γ phases of $CaTiO_3$ for upper and lower spins

According to the results presented in Fig. 3, the density of states due to the lower and upper spins for the α, β and γ-phases of CaTiO$_3$ is the same, i.e., the spin-orbit effects for these systems are hardly noticeable. Moreover, as the phase transformation of the cubic phase into tetragonal and orthorobic, the band gap increases. On the other hand, it can be seen that the tetragonal phase of calcium titanate is characterized by a higher level of the density of electronic states and, accordingly, by the maximum number of vacancies in atomic orbitals.

Further in Table 4 shows and compares the band gap α, β, γ of the phases of CaTiO$_3$ crystals, calculated from the exchange-correlation functional TB-mBJ, since the prediction of such fundamental characteristics of semiconductors has become relevant due to their technological applications, such as optoelectronics and photovoltaics [16]. On the other hand, this is due to the fact that the solid-state computing community pays great attention to the problem of predicting the fundamental band gap of crystals.

**Table 4.** The value of the energy was (in eV) α, β, γ phases of CaTiO$_3$ crystals according to mBJ-calculations

| System | Band gap |
|---|---|
| α-CaTiO$_3$ | 2.83 |
| β-CaTiO$_3$ | 3.07 |
| γ-CaTiO$_3$ | 3.26 |

Figures 4 and 5, respectively, graphs of the absorption capacity and photoconductivity of the α, β, γ phases of CaTiO3 crystals are presented, from which it is easy to see that these crystals have high absorption and optical conductivity coefficients in the high energy region (short-wave radiation). The cubic system of calcium titanate, similar to the small band gap of this phase of the crystal, has the highest photocatalytic activity when illuminated with long-wavelength rays.

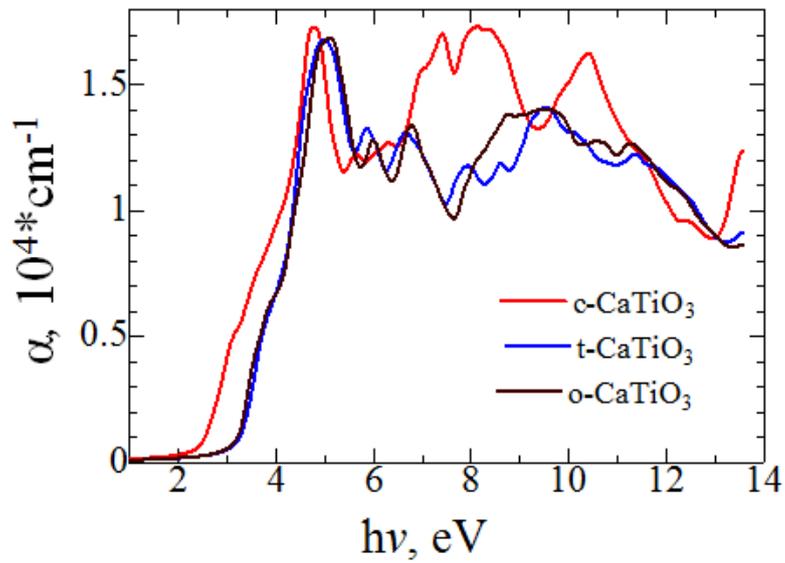

**Figure 4.** Adsorption spectra of α, β, γ phases of CaTiO$_3$ crystals depending on photon energy

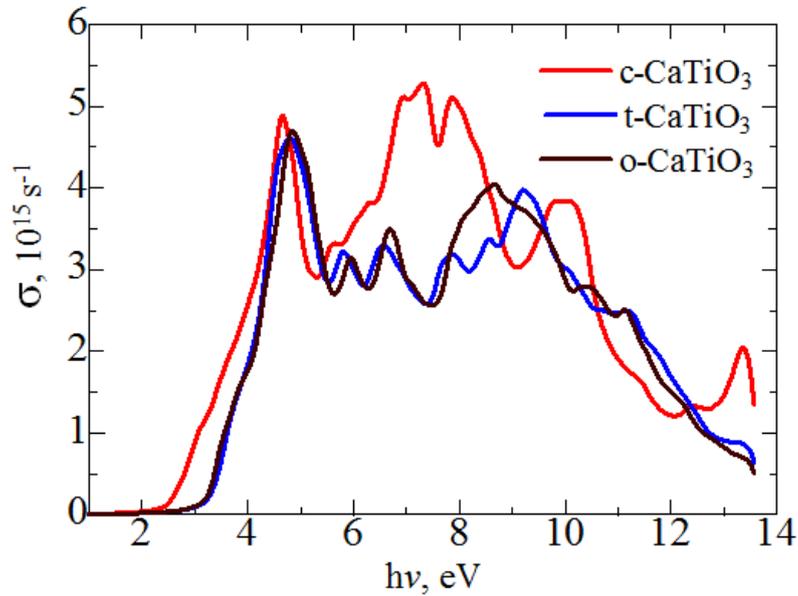

**Figure 5.** Photoconductivity spectra of α, β, γ phases of CaTiO$_3$ crystals depending on photon energy

Similar to the results shown in Figure 4, it can be seen that for the calcium titanate cubic system, photoconductivity starts at the lowest photon energies. In this case, in all ranges of the solar spectrum, the photoconductivity of the cubic phase is higher than that of other structures. This is strongly related to the internal properties of the investigated phases of calcium titanate, especially with the atomic structure, bond length of atoms and their symmetry. It is also particularly related to the bonding dynamics and structural stability of the studied materials that we have explored previously.

Based on the results obtained, it is easy to see that calcium titanate crystals are characterized by large band gaps (an energy barrier for the transition of electrons from the valence band to the conduction band), which limits the efficiency of electron-hole recombination in photoelectric processes. It can be seen that all the studied phases have high coefficients of absorption and optical conductivity in the visible and UV ranges of solar radiation; however, these characteristics are very low in the IR range of light. That is, solar panels based on pure calcium titanate will be ineffective in places where traditional and tandem solar panels are used. On the other hand, they approach the efficient conversion of solar energy in high-altitude places, where the intensity of solar radiation is higher than in other places. However, with the help of modern approaches, for example, doping of transition metals or the influence of pressure, it is possible to control the "tuning" of the band gap of calcium titanate crystals and optimize their optoelectronic properties for photovoltaic applications, since the band gap is the main and one of the most important characteristics of crystals from the point of view of their electrical properties.

The results obtained can be used by other researchers to further study calcium titanate and other similar crystals that are supposed to be synthesized, as well as to determine such important characteristics as "composition-structure-property".

## 4. Conclusions

The The structural characteristics and optoelectronic properties of the α, β, γ phases of $CaTiO_3$ have been studied using quantum chemical calculations. The lattice parameters and volume of the α-, γ-, and β-phases of calcium titanate are obtained and compared with the results of other independent authors. Based on the results of quantum mechanical calculations, the energies of the ground state α, β, γ of the $CaTiO_3$ phases were determined. It has been established that the orthorhombic phase of calcium titanate has the lowest energy and, accordingly, the most stable structure compared to other studied phases. The results obtained showed that the band gap of $CaTiO_3$ increases upon going from α to γ. The calculated optical properties, consisting of absorptivity and photoconductivity, showed that these crystals have high absorptivity and optical conductivity in the high energy region. The high absorption capacity predicts that $CaTiO_3$ is suitable for solar cell applications. Our study provides a theoretical understanding of the justification for the development of high-performance photovoltaic devices based on $CaTiO_3$ and the synthesis of photovoltaic materials with improved performance.

## Funding

The work was supported by the Presidential Fund for Basic Research of the Republic of Tajikistan, grant No. 0122TJ1460.